\documentclass[aps,twocolumn]{revtex4}
\usepackage{amsmath,amssymb}

\begin{document}

\title{Entropic Corrections to Friedmann Equations}
\author{Ahmad  Sheykhi \footnote{
sheykhi@mail.uk.ac.ir}}
\address{Department of Physics, Shahid Bahonar University, P.O. Box 76175, Kerman, Iran\\
         Research Institute for Astronomy and Astrophysics of Maragha (RIAAM), Maragha,
         Iran}

\begin{abstract}
Recently, Verlinde discussed that gravity can be understood as an
entropic force caused by changes in the information associated
with the positions of material bodies. In the Verlinde's argument,
the area law of the black hole entropy plays a crucial role.
However, the entropy-area relation can be modified from the
inclusion of quantum effects, motivated from the loop quantum
gravity. In this note, by employing this modified entropy-area
relation, we derive corrections to Newton's law of gravitation as
well as modified Friedman equations by adopting the viewpoint that
gravity can be emerged as an entropic force. Our study further
supports the universality of the log correction and provides a
strong consistency check on Verlinde's model.
\end{abstract}

 \maketitle

\newpage

It was first pointed out by Jacobson \cite{Jac} that the
hyperbolic second order partial differential Einstein equation has
a predisposition to the first law of thermodynamics. This profound
connection between the first law of thermodynamics and the
gravitational field equations has been extensively observed in
various gravity theories \cite{Elin,Pad}. Recently the study on
the connection between thermodynamics and gravity has been
generalized to the cosmological situations
\cite{Cai2,Cai3,CaiKim,Wang,Cai33,Shey0}, where it was shown that
the differential form of the Friedmann equation on the apparent
horizon in the FRW universe can be rewritten in the form of the
first law of thermodynamics. The extension of this connection has
also been carried out in the braneworld cosmology
\cite{Shey1,Shey2}. The deep connection between the gravitational
equation describing the gravity in the bulk and the first law of
thermodynamics on the apparent horizon reflects some deep ideas of
holography. Although Jacobson's derivation is logically clear and
theoretically sound, the statistical mechanical origin of the
thermodynamic nature of general relativity remains obscure.

Recently, Verlinde \cite{Ver} has put forward an idea similar in
spirit to Jacobson's thermodynamic derivation of the Einstein
equations, where it is argued that Newton's law of gravitation can
be understood as an entropic force caused by information changes
when a material body moves away from the holographic screen.
Quantitatively, when a test particle or excitation moves apart
from the holographic screen, the magnitude of the entropic force
on this body has the form
\begin{equation}\label{F}
F\triangle x=T \triangle S,
\end{equation}
where $\triangle x$ is the displacement of the particle from the
holographic screen, while $T$ and $\triangle S$ are the
temperature and the entropy change on the screen, respectively.
Verlinde's derivation of Newton's law of gravitation at the very
least offers a strong analogy with a well understood statistical
mechanism. Therefore, this derivation opens a new window to
understand gravity from the first principles. The study on the
entropic  force has raised a lot of enthusiasm recently (see
\cite{Cai4,Other, Modesto,Yi} and references therein).

It is important to note that in Verlinde discussion, the black
hole entropy $S$ plays a crucial role in the derivation of
Newton's law of gravitation. Indeed, the derivation of Newton's
law of gravity depends on the entropy-area relationship $S=A/4\ell
_p^2$ of black holes in Einstein's gravity, where $A =4\pi R^2$
represents the area of the horizon and $\ell _p^2=G\hbar/c^3$ is
the Planck length. However, this definition can be modified from
the inclusion of quantum effects, motivated from the loop quantum
gravity (LQG). The quantum corrections provided to the
entropy-area relationship leads to the curvature correction in the
Einstein-Hilbert action and vice versa \cite{Zhu,Suj}. The
corrected entropy takes the form \cite{Zhang}
\begin{equation}\label{S}
S=\frac{A}{4\ell _p^2}-\beta \ln {\frac{A}{4\ell _p^2}}+\gamma
\frac{\ell _p^2}{A}+\mathrm{const},
\end{equation}
where $\beta$ and $\gamma$ are dimensionless constants of order
unity. The exact values of these constants are not yet determined
and still an open issue in loop quantum cosmology. These
corrections arise in the black hole entropy in LQG due to thermal
equilibrium fluctuations and quantum fluctuations \cite{Rovelli}.
Taking  the corrected entropy-area relation (\ref{S}) into
account, we will derive the corrections to the Newton's law of
gravitation as well as the modified Friedman equations. We rewrite
Eq. (2) in the following form
\begin{equation}
\label{S2}
 S=\frac{A}{4\ell _p^2}+{s}(A),
\end{equation}
where $s(A)$ stands for the correction terms in the entropy
expression. We adopt the viewpoint of \cite{Ver}. Suppose we have
two masses one a test mass and the other considered as the source
with respective masses $m$ and $M$. Centered around the source
mass $M$, is a spherically symmetric surface $\mathcal {S}$ which
will be defined with certain properties that will be made explicit
later. To derive the entropic law, the surface $\mathcal {S}$ is
between the test mass and the source mass, but the test mass is
assumed to be very close to the surface as compared to its reduced
Compton wavelength $\lambda_m=\frac{\hbar}{mc}$. When a test mass
$m$ is a distance $\triangle x = \eta \lambda_m$ away from the
surface $\mathcal {S}$, the entropy of the surface changes by one
fundamental unit $\triangle S$ fixed by the discrete spectrum of
the area of the surface via the relation
\begin{equation}
\label{S3}
 \triangle S=\frac{\partial S}{\partial A}\triangle A=\left(\frac{1}{4\ell _p^2}+\frac{\partial{s}(A)}{\partial A}\right)\triangle
 A.
\end{equation}
The energy of the surface $\mathcal {S}$ is identified with the
relativistic rest mass of the source mass:
\begin{equation}
\label{Ec} E=M c^2.
\end{equation}
On the surface $\mathcal {S}$, there live a set of ``bytes" of
information that scale proportional to the area of the surface so
that
\begin{equation}
\label{AQN}
 A=QN,
 \end{equation}
where $N$ represents the number of bytes and $Q$ is a fundamental
constant which should be specified later. Assuming the temperature
on the surface is $T$, and then according to the equipartition law
of energy \cite{Pad1}, the total energy on the surface is
\begin{equation}
\label{E}
 E=\frac{1}{2}Nk_B T.
 \end{equation}
Finally, we assume that the force on the particle follows from the
generic form of the entropic force governed by the thermodynamic
equation of state
\begin{equation}\label{F2}
F=T \frac{\triangle S}{\triangle x},
\end{equation}
where $\triangle S$ is one fundamental unit of entropy when
$|\triangle x|= \eta \lambda_m$, and the entropy gradient points
radially from the outside of the surface to inside. Note that $N$
is the number of bytes and thus $\triangle N=1$, hence from
(\ref{AQN}) we have $\triangle A=Q$. Now, we are in a position to
derive the entropic-corrected Newton's law of gravity. Combining
Eqs. (\ref{S3})- (\ref{F2}), we easily obtain
\begin{equation}\label{F3}
F=-\frac{Mm}{R^2}\left(\frac{Qc^3}{2\pi k_B \hbar
\eta}\right)\left[\frac{1}{4\ell _p^2}+\frac{\partial{s}}{\partial
A}\right]_{A=4\pi R^2},
\end{equation}
This is nothing but the Newton's law of gravitation to the first
order provided we define $Q=8\pi k_B \eta \ell_p^4$. Thus we reach
\begin{equation}\label{F4}
F=-\frac{GMm}{R^2}\left[1+4\ell_p^2 \frac{\partial{s}}{\partial
A}\right]_{A=4\pi R^2},
\end{equation}
Finally, using Eq. (\ref{S}) we obtain the modified Newton's law
of gravitation as
\begin{equation}\label{F5}
F=-\frac{GMm}{R^2}\left[1-\frac{\beta}{\pi}\frac{\ell_p^2}{R^2}
-\frac{\gamma}{4\pi^2}\frac{\ell_p^4}{R^4}\right],
\end{equation}
Thus, with the  correction in the entropy expression, we see that
the Newton's law will modified accordingly. As we mentioned, these
corrections are well motivated from bottom-up quantum gravity
theories. The log correction to the area-entropy relation appears
to have an almost universal status, having been derived from
multiple different approaches to the calculation of entropy from
counting microscopic states in different quantum gravity models.
Since the last two terms in Eq. (\ref{F5}) can be comparable to
the first term only when $R$ is very small, the corrections make
sense only at the very small distances. When $R$ becomes large,
the entropy-corrected Newton's law reduces to the usual Newton's
law of gravitation.

Let us compare our result with that obtained in \cite{Modesto}.
The first correction term originates from the log correction in
Eq. (17) of \cite{Modesto} is similar to ones we obtained in Eq.
(\ref{F5}), however, it seems that the second correction term in
Eq. (17) of \cite{Modesto} is not reasonable. Physically, the
effect of the correction terms on the quantity should be less than
the uncorrected quantity. Similarly, the contribution of the first
correction term in the physical quantity (force here) should be
more from the second term and so on. For all above reasons, we
think the second correction term in Eq. (17) of \cite{Modesto} is
not correct, and the corrected form is that presented in our note.
The origin of this difference is due to the fact that the second
(volume) correction term to the entropy expression in
\cite{Modesto} is not indeed a correction term, although it was
argued by \cite{Modesto} that this term is also emerged in a model
for the microscopic degrees comprising the black hole entropy in
LQG \cite{smolin}. Let us stress here that although in the
literature there is doubt about the second correction term in
entropy-corrected relation, however, it is  widely believed \cite
{Zhang} that the next quantum correction term to black hole
entropy have the form $\ell_p^2/A$, which leads to the resonable
correction terms to Newton's law of gravitation (\ref{F5}) as we
have shown in the present work and will lead to corrected modified
Friedmann equation as we will see later.

Next, we extend our discussion to the cosmological setup. Assuming
the background spacetime to be spatially homogeneous and isotropic
which is given by the Friedmann-Robertson-Walker (FRW) metric
\begin{equation}
ds^2={h}_{\mu \nu}dx^{\mu} dx^{\nu}+R^2(d\theta^2+\sin^2\theta
d\phi^2),
\end{equation}
where $R=a(t)r$, $x^0=t, x^1=r$, the two dimensional metric
$h_{\mu \nu}$=diag $(-1, a^2/(1-kr^2))$. Here $k$ denotes the
curvature of space with $k = 0, 1, -1$ corresponding to open,
flat, and closed universes, respectively. The dynamical apparent
horizon, a marginally trapped surface with vanishing expansion, is
determined by the relation $h^{\mu
\nu}\partial_{\mu}R\partial_{\nu}R=0$. A simple calculation gives
the apparent horizon radius for the FRW universe
\begin{equation}
\label{radius}
 R=ar=\frac{1}{\sqrt{H^2+k/a^2}}.
\end{equation}
We also assume the matter source in the FRW universe is a perfect
fluid with stress-energy tensor
\begin{equation}\label{T}
T_{\mu\nu}=(\rho+p)u_{\mu}u_{\nu}+pg_{\mu\nu}.
\end{equation}
Due to the pressure, the total mass $M = \rho V$ in the region
enclosed by the boundary $\mathcal S$ is no longer conserved, the
change in the total mass is equal to the work made by the pressure
$dM = -pdV$ , which leads to the well-known continuity equation
\begin{equation}\label{Cont}
\dot{\rho}+3H(\rho+p)=0,
\end{equation}
where $H=\dot{a}/a$ is the Hubble parameter. It is instructive to
first derive the dynamical equation for Newtonian cosmology.
Consider a compact spatial region $V$ with a compact boundary
$\mathcal S$, which is a sphere with physical radius $R= a(t)r$.
Note that here $r$ is a dimensionless quantity which remains
constant for any cosmological object partaking in free cosmic
expansion. Combining the second law of Newton for the test
particle $m$ near the surface, with gravitational force (\ref{F5})
we get
\begin{equation}\label{F6}
F=m\ddot{R}=m\ddot{a}r=-\frac{GMm}{R^2}\left[1-\frac{\beta}{\pi}\frac{\ell_p^2}{R^2}
-\frac{\gamma}{4\pi^2}\frac{\ell_p^4}{R^4}\right],
\end{equation}
We also assume $\rho=M/V$ is the energy density of the matter
inside the the volume $V=\frac{4}{3} \pi a^3 r^3$. Thus, Eq.
(\ref{F6}) can be rewritten as
\begin{equation}\label{F7}
\frac{\ddot{a}}{a}=-\frac{4\pi
G}{3}\rho\left[1-\frac{\beta}{\pi}\frac{\ell_p^2}{R^2}
-\frac{\gamma}{4\pi^2}\frac{\ell_p^4}{R^4}\right],
\end{equation}
This is nothing but the entropy-corrected dynamical equation for
Newtonian cosmology. The main difference between this equation and
the standard dynamical equation for Newtonian cosmology is that
the correction terms now depends explicitly on the radius $R$.
However, we can remove this confusion. Assuming that for Newtonian
cosmology the spacetime is Minkowskian with $k=0$, then we get
$R=1/H$, and we can rewrite Eq. (\ref{F7}) in the form
\begin{equation}\label{F8}
\frac{\ddot{a}}{a}=-\frac{4\pi G}{3}\rho\left[1-\frac{\beta
\ell_p^2}{\pi}\left(\frac{\dot{a} }{a}\right)^2 -\frac{\gamma
\ell_p^4}{4\pi^4}\left(\frac{ \dot{a}}{a}\right)^4\right].
\end{equation}
It was argued in \cite{Cai4} that for deriving the Friedmann
equations of FRW universe in general relativity, the quantity that
produces the acceleration is the active gravitational mass
$\mathcal M$ \cite{Pad2}, rather than the total mass $M$ in the
spatial region $V$. With the entropic corrections terms, the
active gravitational mass $\mathcal M$ will also modified as well.
On one side, from Eq. (\ref{F7}) with replacing $M$ with $\mathcal
M$ we have
\begin{equation}\label{M1}
\mathcal M =-\frac{\ddot{a}
a^2}{G}r^3\left[1-\frac{\beta}{\pi}\frac{\ell_p^2}{R^2}
-\frac{\gamma}{4\pi^2}\frac{\ell_p^4}{R^4}\right]^{-1}.
\end{equation}
On the other side, the active gravitational mass is  defined as
\cite{Cai4}
\begin{equation}\label{Int}
\mathcal M =2
\int_V{dV\left(T_{\mu\nu}-\frac{1}{2}Tg_{\mu\nu}\right)u^{\mu}u^{\nu}}.
\end{equation}
A simple calculation leads
\begin{equation}\label{M2}
\mathcal M =(\rho+3p)\frac{4\pi}{3}a^3 r^3.
\end{equation}
Equating Eqs. (\ref{M1}) and (\ref{M2}) we  find
\begin{equation}\label{addot}
\frac{\ddot{a}}{a} =-\frac{4\pi
G}{3}(\rho+3p)\left[1-\frac{\beta}{\pi}\frac{\ell_p^2}{R^2}
-\frac{\gamma}{4\pi^2}\frac{\ell_p^4}{R^4}\right].
\end{equation}
This is the modified acceleration equation for the dynamical
evolution of the  FRW universe. Multiplying $\dot{a}a$ on both
sides of Eq. (\ref{addot}), and using the continuity equation
(\ref{Cont}), after integrating we find
\begin{equation}\label{Fried1}
H^2+\frac{k}{a^2} =\frac{8\pi
G}{3}\rho\left[1-\frac{\beta}{\pi}\frac{\ell_p^2}{\rho
R^2}\int{\frac{d(\rho a^2)}{a^2}} -\frac{\gamma}{4\pi^2}\frac{a^2
\ell_p^4}{\rho R^4} \int{\frac{d(\rho a^2)}{a^4}}\right].
\end{equation}
Now, in order to calculate the integrations in the correction
terms we need to find $\rho=\rho(a)$. Assume the equation of state
parameter $w=p/\rho$ is a constant, the continuity equation
(\ref{Cont}) can be integrated immediately to give
\begin{equation}\label{rho}
\rho=\rho_0 a^{-3(1+w)},
\end{equation}
where $\rho_0$, an integration constant, is the present value of
the energy density. Inserting relation (\ref{rho}) in Eq.
(\ref{Fried1}), after integration, we obtain
\begin{equation}\label{Fried2}
H^2+\frac{k}{a^2} =\frac{8\pi G}{3}\rho\left[1-\frac{\beta
(1+3w)}{3\pi(1+w)}\frac{\ell_p^2}{R^2} -\frac{\gamma
(1+3w)}{4\pi^2 (5+3w)}\frac{\ell_p^4 }{R^4} \right].
\end{equation}
Using Eq. (\ref{radius}) we can  further rewrite the above
equation as
\begin{eqnarray}\label{Fried3}
&&\left(H^2+\frac{k}{a^2}\right)\left[1-\frac{\beta \ell_p^2
(1+3w)}{3\pi(1+w)}\left(H^2+\frac{k}{a^2}\right) \right. \nonumber
\\
&& \left.-\frac{\gamma \ell_p^4 (1+3w)}{4\pi^2 (5+3w)}
\left(H^2+\frac{k}{a^2}\right)^{2}\right]^{-1}=\frac{8\pi
G}{3}\rho.
\end{eqnarray}
If $\beta$ and $\gamma$ are  viewed as small quantities, then the
above equation can be expanded up to the linear order of $\beta$
and $\gamma$. The result is
\begin{eqnarray}\label{Fried4}
&&\left(H^2+\frac{k}{a^2}\right)+\frac{\beta \ell_p^2
(1+3w)}{3\pi(1+w)}\left(H^2+\frac{k}{a^2}\right)^{2}  \nonumber
\\
&& +\frac{\gamma \ell_p^4 (1+3w)}{4\pi^2 (5+3w)}
\left(H^2+\frac{k}{a^2}\right)^{3}=\frac{8\pi G}{3}\rho,
\end{eqnarray}
which is in complete agreement with the result of \cite{Cai33}
(see also \cite{Shey0}). In this way we derive the
entropy-corrected Friedmann equation of FRW universe by
considering gravity as an entropic force caused by changes in the
information associated with the positions of material bodies. In
the absence of the correction terms $(\beta=0=\gamma)$, one
recovers the well-known Friedmann equation in standard cosmology.
Since the last two terms in Eq. (\ref{Fried2}) can be comparable
to the first term only when $a$ is very small, the corrections
make sense only at early stage of the universe where $a\rightarrow
0$. When the universe becomes large, the entropy-corrected
Friedmann equation reduces to the standard Friedman equation.

In summary, we have shown that with the entropy corrections to the
area-relation of the black hole entropy, the  Newton's law of
gravitation and the Friedmann equations will be modified
accordingly. These corrections are motivated from the LQG which is
one of the promising theories of quantum gravity.  We derived the
correction terms to the Newton's law  of gravity as well as
modified Friedmann equations of the FRW universe starting from the
holographic principle and the equipartition law of energy by using
Verlinde's argument that gravity appears as an entropic force. In
particular, we have found that an apparently universal log
correction to the area-entropy, yields deviations from Newton's
law and Friedman equations that are identical in form to those
obtained from perturbative quantum gravity. This at once sheds
light on the reason for the universality of the log correction and
provides a strong consistency check on Verlinde's model.

%%%%%%%%%%%%%%%%%%%%%%%%%%%%%%%%%%%%%%%%%%%%%%%%%%%%%%%%%%%%%%%%%%%%%%%
\acknowledgments{I am grateful to Prof. B. Wang and Prof. R.G. Cai
for helpful discussions and reading the manuscript. This work has
been supported by Research Institute for Astronomy and
Astrophysics of Maragha.}
%%%%%%%%%%%%%%%%%%%%%%%%%%%%%%%%%%%%%%%%%%%%%%%%%%%%%%%%%%%%%%%%%%%%%%%%%%%

\end{document}